\documentclass[pra,twocolumn,showpacs,amsmath,amssymb,superscriptaddress,longbibliography]{revtex4-1}
\usepackage[none]{hyphenat}
\usepackage{amsmath, amsfonts,amssymb}
\usepackage{comment}
\usepackage{soul}
\usepackage{dcolumn}% Align table columns on decimal point
\usepackage{mathtools}			 % Math capabilities
\usepackage{physics}
\usepackage{graphicx}
\usepackage{epsfig}
\usepackage{comment}
\usepackage{color}
\usepackage{textcomp}
\usepackage[colorlinks,citecolor=blue]{hyperref}

\begin{document}
\preprint{APS/123-QED}

\title{Non-Markovian Collective Emission of Giant emitters in the Zeno Regime}
%\title{Zeno Regime of Collective Emission in an Array of \boldmath$N$ Giant Emitters}
\author{Qing-Yang Qiu}
\affiliation{School of Physics and Institute for Quantum Science and Engineering, Huazhong University of Science and Technology, Wuhan, 430074, P. R. China}
\affiliation{Wuhan institute of quantum technology, Wuhan, 430074, China}

\author{Xin-You L\"{u}}\email{xinyoulu@hust.edu.cn}
\affiliation{School of Physics and Institute for Quantum Science and Engineering, Huazhong University of Science and Technology, Wuhan, 430074, P. R. China}
\affiliation{Wuhan institute of quantum technology, Wuhan, 430074, China}

\date{\today}% It is always \today, today,
             %  but any date may be explicitly specified
\begin{abstract}
  We explore the collective Zeno dynamics of giant artificial atoms that are coupled, via multiple coupling points, to a common photonic or acoustic reservoir. In this regime, the establishment of atomic cooperativity and the revivification of exponential decay are highly intertwined, which is utterly beyond the non-Markovian regime with only retarded backaction. We reveal that giant atoms build up their collective emission smoothly from the decay rate of zero to that predicted by Markovian approximation and show great disparity between different waveguide QED setups. As a comparison, the steplike growth of instantaneous decay rates in the retardation-only picture has also been shown. All of these theoretical pictures predict the same collective behavior in the long time limit. From a phenomenological standpoint, we observe that the atomic superradiance exhibits significant directional property.  In addition, the subradiant photons feature prolonged oscillation in the early stage of collective radiance, where the energy is exchanged remarkably between giant emitters and the field. Our results might be probed in state-of-art waveguide QED experiments and fundamentally broaden the fields of collective emission in systems with giant atoms.
\end{abstract}

\maketitle
\section{INTRODUCTION}\label{Sec I}
Spontaneous emission from an unstable system, as one of the earliest quantum field theory phenomena, was first predicted by Einstein and subsequently extended to multiatomic ensembles in $1953$, which is known as Dicke
superradiance\,\cite{Dicke1954}. These significant contributions have profoundly solidified the concept of exponential decay law. However, it is not the whole physical picture because the realistic spontaneous emission is generally exposed to a non-Markovian environment. Indeed, non-Markovianity can have a wealth of  physical origins\,\cite{PhysRevD.45.2843,PhysRevD.47.1576,PhysRevA.87.013820,PhysRevA.89.022109,Groblacher2015,RevModPhys.88.021002,RevModPhys.89.015001} with a hallmark of strong informational backflow. Light-matter interactions within such a non-Markovian bath open new routes for unprecedented  phenomena, such as genuinely non-Markovian steady states\,\cite{Andreas2022,cilluffo2024}, collective superradiant burst\,\cite{PhysRevLett.124.043603,PhysRevA.107.023723} and  engineered Floquet bound state\,\cite{PhysRevLett.131.050801}. In particular, the retardation induced non-Markovianity has been demonstrated to endow collective radiation with a common intuition:  each emitter initially undergoes independent exponential decay, altering its decay behavior upon interaction with delayed photons emitted by other emitters\,\cite{PhysRevLett.124.043603}.

 Recently, Zeno regime of spontaneous emission comes to the fore in the community of waveguide quantum electrodynamics (QED), revealing how pointlike atoms cooperatively modify their decay behavior in the building of collective emission\,\cite{Zhang2023}. In this context, atoms complete their significant transition from nonexponential to exponential decay by exchanging virtual photons in a very short time. This time scale evaluates the duration of non-exponential decay, i.e., the Zeno time $\tau_{Z}$, during which the atomic survival probability is quadratic in time\,\cite{Wilkinson1997,Crespi2019,Giacosa2021,Liu2023}.

\begin{figure*}
  \centering
  % Requires \usepackage{graphicx}
  \includegraphics[width=18cm]{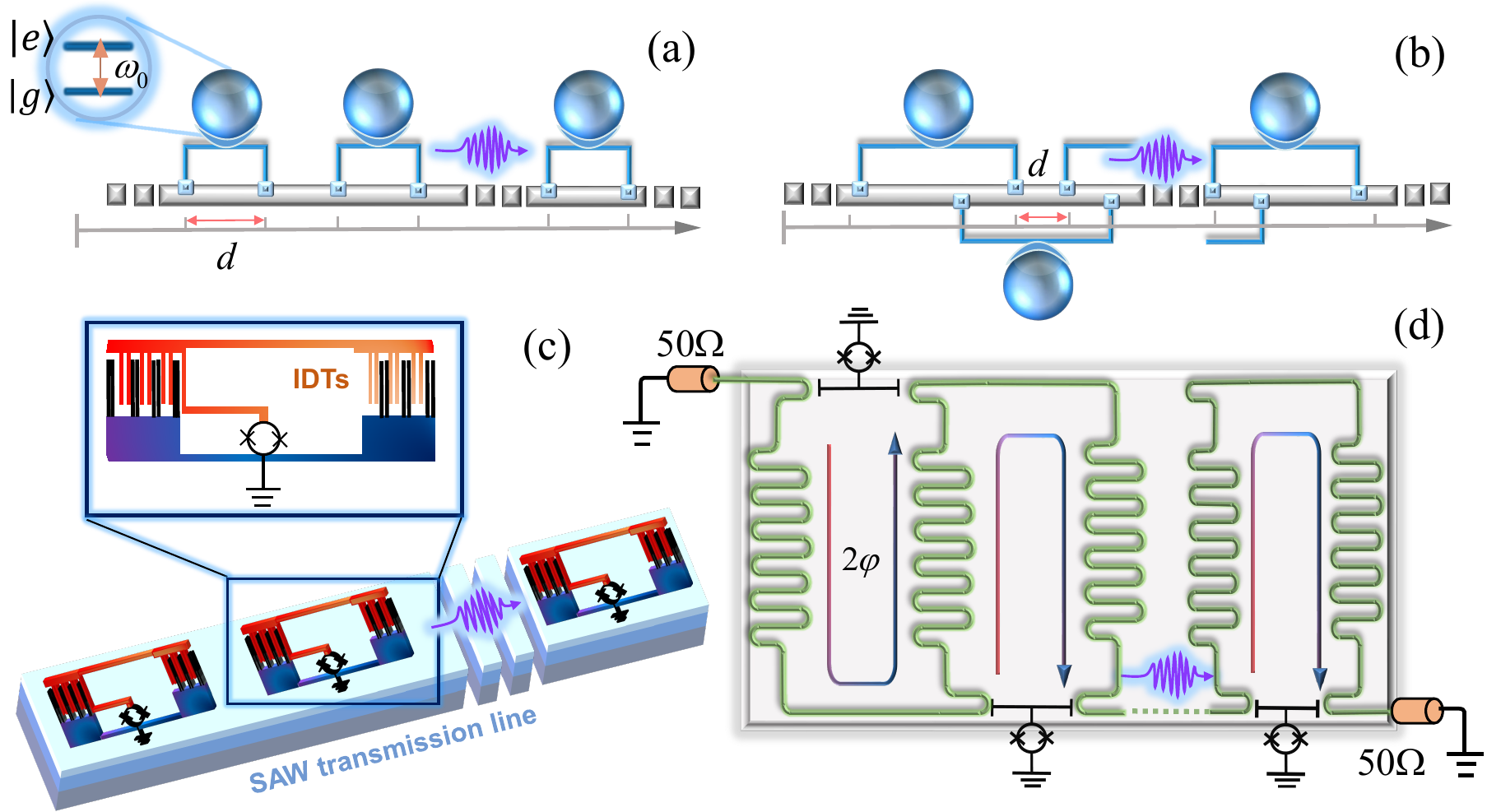}
  \caption{Schematic diagram of theoretical model and experimental implementation for a linear chain of $N$ GAs. Panels (a) and (b) respectively show $N$ two-legged separate and braided superconducting qubits are coupled to a 1D waveguide at multiple connecting points with the light blue dots denoting the coupling points. The experimental setup in a separate case is depicted accordingly in (c), i.e., $N$ transmon qubits (superconducting quantum interference devices marked by black circles) are embedded in a surface acoustic waves transmission line. Each qubit interacts with the propagating phonons via piezoelectric effect utilizing the large interdigital transducers from its two islands. For the braided case, panel (d) exhibits $N$ X-mon qubits coupled capacitively to a meandering microwave coplanar waveguide with the central waveguide (dark green line) terminated to $50\Omega$.}\label{fig1}
\end{figure*}

Featuring exotic self-interference effects and the nature of non-local couplings\,\cite{PhysRevA.90.013837,PhysRevA.95.053821,Frisk_Kockum_2020}, ``giant'' artificial atoms (GAs)  have spurred a rapidly growing interest in the emerging area of waveguide QED. It has been manipulated to interact with surface acoustic waves (SAWs)\,\cite{Andersson2015} or the meandering coplanar waveguide (CPW) \,\cite{Kannan2020,PhysRevA.103.023710,PhysRevX.13.021039} via multiple connecting points as sketched in Fig.\ref{fig1}. Leveraging the state-of-the-art quantum device, such a novel designation has recently been extended to ``giant" ferromagnetic spin ensemble\,\cite{Wang2022,PhysRevA.108.063715}. Of particular interest is the exploration of novel non-Markovian effects with GAs and considerable effort has been made to manifest a rich variety of  intriguing outcomes, including  tunable localization-delocalization quantum phase transition\,\cite{PhysRevA.106.063717}, oscillating bound states\,\cite{PhysRevResearch.2.043014,PhysRevA.107.023716, Xu_2024}, enhanced Dicke superradiance\,\cite{Qiu2023} and so on\,\cite{PhysRevA.106.063703,PhysRevResearch.4.023198,PhysRevA.106.033522,PhysRevA.109.033711,PhysRevA.109.023712}. However, it seems that the mechanism of non-negligible time-delayed feedback underlies all of these efforts. Thus, the scenario of collective emission beyond the retardation, i.e., GAs to Zeno region, is still unexplored.

In this work, we have investigated the collective radiative dynamics of a linear chain of $N$ GAs coupling to a 1D waveguide and show that it exhibits prominent non-Markovianity beyond retardation. To incorporate the relaxation dynamics into the Zeno regime, the array of GAs is designed to guarantee $\tau_{0}\ll \tau_{Z}$, where $\tau_{0}$ is the time taken by a single photon or phonon traveling between adjacent coupling points. On this occasion, the GAs build up their cooperativity obeying the combined action of time-delayed feedback and memory effect of the electromagnetic field. As a consequence, we find that the non-Markovianity can be enhanced by increasing the coupling points or modifying the waveguide QED setups. In contrast to the previous achievements obtained outside the Zeno region, the full development of collective emission shows a dominating signature: each emitter grows their instantaneous decay rate smoothly from zero to that predicted by Markovian theory and predictions from different waveguide QED setups are distinguished by the building speed of emission and the magnitude of the memory effect. We also put forward steplike growth of decay rates in the  retardation-only picture for the purpose of comparison. In an anticipated manner, these predictions can eventually coalesce with each other at long enough time. To get further insight into the collective Zeno physics, we have also studied the dynamics of the field emitted from the array of  GAs, during the establishment of collective radiation. Interestingly, we find that the chiral radiation occurs when the superradiant development of GAs is asymmetric. Moreover, a remarkable oscillation can be captured in the early stage of  subradiance, where the atoms are periodically emitting and absorbing photons.
\section{MODEL HAMILTONIAN AND ZENO TIME OF WAVEGUIDE QED}\label{Sec II}
We consider $N$ GAs interacting with an open 1D waveguide through multiple coupling points and apply a strict two-level approximation with the ground state $\ket{g}$ and the excited state $\ket{e}$. Figures. \ref{fig1} (a) and \ref{fig1} (b) illustrate the theoretical model of two-legged separate (all coupling points of each atom are outside those of another atom) and braided (part of the coupling points of each atom is inside those of another atom) GAs, respectively.  The waveguide QED setups with giant atoms have at least two different experimental platforms associated with SAWs or superconducting transmission line, as sketched respectively in Figs. \ref{fig1} (c) and \ref{fig1} (d).

The total Hamiltonian for the GAs+bath system reads $H_{{\rm{tot}}}=H_{A}+H_{B}+H_{{\rm{int}}}$, where $H_{A}=\omega_{0}\sum\limits_{n=1}^{N}\sigma_{n}^{\dagger}\sigma_{n}$ is the free Hamiltonian for the GAs of resonance frequency $\omega_{0}$, with the creation operator of the $n^{{\rm{th}}}$ atom being $\sigma^{\dagger}_{n}=\ket{e}_{n}\bra{g}$;  $H_{B}=\int_{-\Lambda}^{\Lambda}\frac{dk}{2\pi}\omega_{k}a_{k}^{\dagger}a_{k}$ is the free Hamiltonian for the 1D continuum of bosonic modes where $a_{k}^{\dagger}$ is the generation operator of the
waveguide mode with wave number $k$, satisfying bosonic commutation relation $[a_{k},a_{k'}^{\dagger}]=2\pi\delta(k-k^{\prime})$. Here we adopt a linear dispersion relation $\omega_{k}=|k|v_{g}$ with $v_{g}$ being the group velocity of the field. To isolate the non-Markovian effect stemming from the band edges of the photonic or phononic bath\,\cite{PhysRevA.58.4168,PhysRevX.11.041043,PhysRevA.105.062207}, the resonant wave number $k_{0}=\omega_{0}/v_{g}$ is assumed to be far enough away from the cutoff of  the wave number $\Lambda$. The nonlocal interactions between $M$-legged GAs and bosonic field are described by ($\hbar=1$)
\begin{align}\label{eq1}
\!\!H_{{\rm{int}}}\!=\!-i\sum_{n=1}^{N}\sum_{m=1}^{M}\!\int_{-\Lambda}^{\Lambda}\frac{dk}{2\pi}g_{k}(\sigma_{n}\!+\!\sigma^{\dagger}_{n})a_{k}e^{ikx^{n}_{m}}\!+\!{\rm{H.c.}},
\end{align}
where $x^{n}_{m}$ denotes the $m^{{\rm{th}}}$ coupling point of  the $n^{{\rm{th}}}$  giant atom ($x^{n}_{m_{1}}<x^{n}_{m_{2}}$  if $m_{1}<m_{2}$). In the context of waveguide QED, there are many general choices for the specific form of coupling strength $g_{k}$, among which constant and linear spectral density are frequently considered.  The above two types of  coupling formalism are referred to as ``const-wQED" and ``lin-wQED" corresponding to $g_{k}=\sqrt{\Gamma_{0}v_{g}/2}$ and $g_{k}=\sqrt{\Gamma_{0}v_{g}|k|/2k_{0}}$, respectively, with $\Gamma_{0}$ being the relaxation rate at a single coupling point.
\begin{figure}
  \centering
  % Requires \usepackage{graphicx}
  \includegraphics[width=8.6cm]{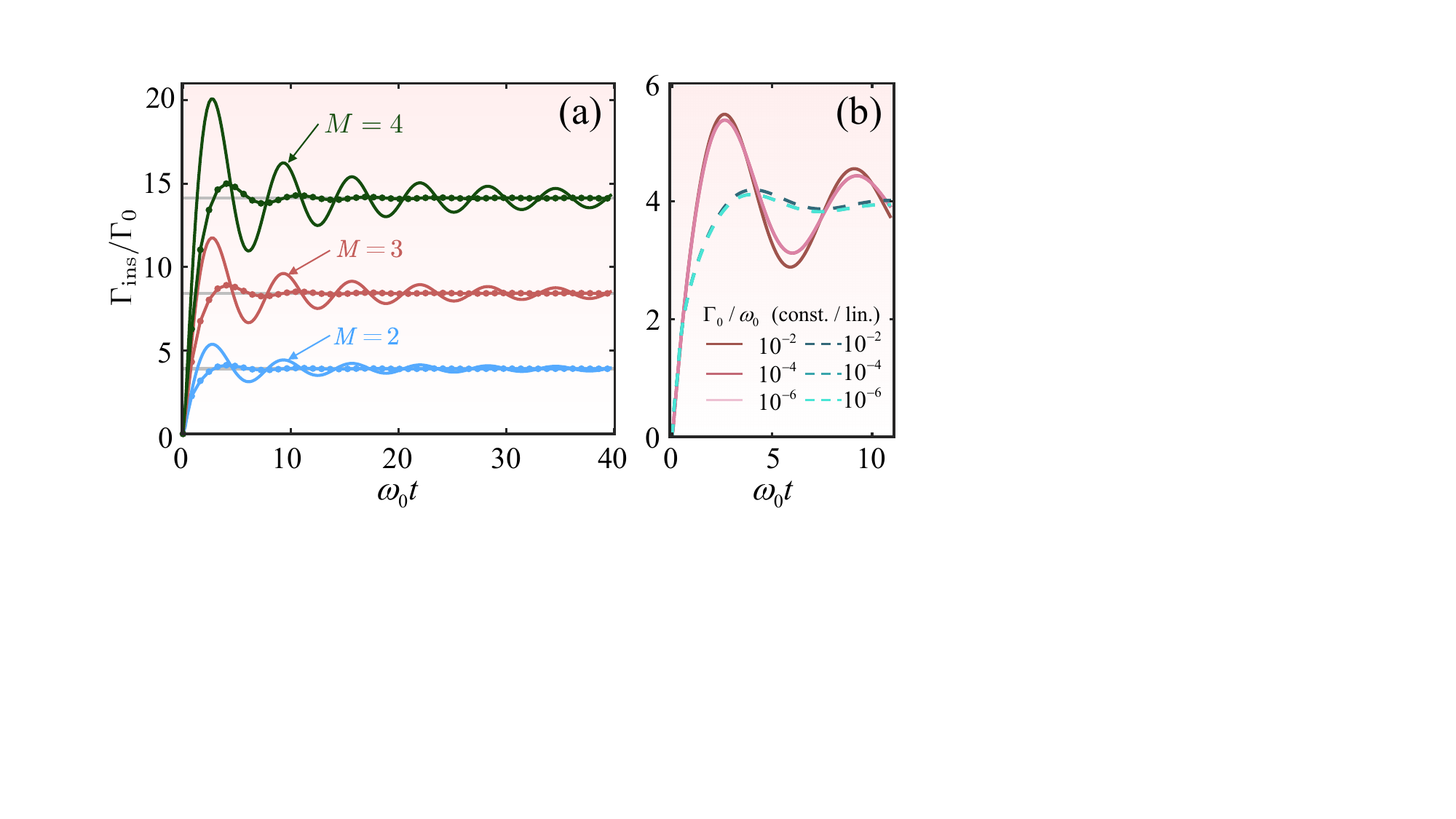}
  \caption{The instantaneous decay rate $\Gamma_{{\rm{ins}}}$ (in units of $\Gamma_{0}$) of a single giant atom as a function of scaled time $\omega_{0}t$. Time evolved decay rates are depicted by (a) gradually increasing the atomic coupling points with fixed $\Gamma_{0}/\omega_{0}=10^{-4}$ or (b) changing the ratio of $\Gamma_{0}/\omega_{0}$.  In panel (a), we distinguish different coupling points by coloring: $M$=2 (blue), $M$=3 (red), $M$=4 (green), and the numerical results obtained from const-wQED (solid lines) or lin-wQED (dotted lines) have been marked.  Also, we utilize gradient color to distinguish the data under different ratios in panel (b), with the predictions of  const-wQED (solid lines) or lin-wQED (dashed lines) have also been marked. Other parameters are $\Lambda/\omega_{0}=10^{4}$ and $d=0.1\pi/\omega_{0}$. }\label{fig2}
\end{figure}
Commonly, the Hamiltonian $H_{{\rm{tot}}}$ can not be solved exactly and the rotating-wave approximation (RWA) is usually applied. Instead, the counterrotating terms in Eq.(\ref{eq1}) can not be neglected in our case, since the time scale we focused on is extremely short. In order to analyze the impact of the full Hamiltonian \ref{eq1} on the short-time collective dynamics, we employ a polaronlike unitary transform\,\cite{PhysRevA.35.4253,RevModPhys.94.045003,PhysRevA.108.043712} on  $H_{{\rm{tot}}}$, i.e., $H_{{\rm{eff}}}=e^{iU}H_{{\rm{tot}}}e^{-iU}$ with
\begin{align}\label{eq2}
\!\!U\!=\!\sum_{n=1}^{N}\sum_{m=1}^{M}\!\int_{-\Lambda}^{\Lambda}\frac{dk}{2\pi}\frac{g_{k}}{\omega_{0}+\omega_{k}}\!(\sigma_{n}\!+\!\sigma^{\dagger}_{n})a_{k}e^{ikx^{n}_{m}}\!+\!{\rm{H.c.}}.
\end{align}
The transformed Hamiltonian is given by $H_{{\rm{eff}}}=H_{A}+H_{B}+H_{{\rm{JC}}}+\mathcal{O}(g^{2}_{k})$ with
\begin{align}\label{eq3}
H_{{\rm{JC}}}=-i\sum_{n=1}^{N}\sum_{m=1}^{M}\int_{-\Lambda}^{\Lambda}\frac{dk}{2\pi}g^{{\rm{JC}}}_{k}\sigma_{n}^{\dagger}a_{k}e^{ikx^{n}_{m}}+\!{\rm{H.c.}},
\end{align}
where $g^{{\rm{JC}}}_{k}=\frac{2\omega_{0}g_{k}}{\omega_{0}+\omega_{k}}$ is the scaled coupling of  the Jaynes- Cummings (JC) type interaction; $\mathcal{O}(g^{2}_{k})$ contains terms of order higher than $g_{k}$  and thus can be safely dropped in the limit of $g_{k}\ll 1$.

We now proceed by determining the duration of nonexponential decay for a single excited giant atom with $M$ legs, which can be well characterized by the Zeno time $\tau_{Z}$.  After some standard algebra, we obtain the Zeno time $\tau_{Z}^{-2}=\frac{2\Gamma_{0}\omega_{0}}{\pi}M\ln(\Lambda/k_{0})$ for the lin-wQED  and the one for the const-wQED has the form of (see Appendix. \ref{A} for more details)
\begin{align}\label{eq4}
\tau_{Z}^{-2}\!=\!\frac{2\Gamma_{0}\omega_{0}}{\pi}[M^{2}+2\sum_{n=1}^{M-1}n\varphi(\chi_{n}(1+\Lambda)\!-\!\chi_{n}(1))],
\end{align}
where $\chi_{n}(x)\!\equiv\!(M-n)[\sin(n\varphi){ {\rm{Ci}}(n\varphi x)}\!-\!\cos(n\varphi){{\rm{Si}}(n\varphi x)}]$ and $\varphi\equiv k_{0}d$ have been introduced for simplicity, with $d$ the minimal distance between adjacent coupling points. Note that the distance between two coupling points from a single giant atom is $d$ for separate two-legged GAs and $3d$ for braided two-legged GAs, as we can see from Figs.\ref{fig1}(a)-(b). The integral functions ${\rm{Ci}}(x)$ and ${\rm{Si}}(x)$ are defined as follows:
\begin{align}\label{eq5}
{\rm{Csi}}(x)\equiv {\rm{Ci}}(x)+i{\rm{Si}}(x)\equiv \int_{x}^{\infty}dz\frac{e^{iz}}{z}.
\end{align}

We note that the results of Zeno time for the two mentioned waveguide-QED setups are cutoffdependent, arising from the self-interference effect of an individual giant atom, which is different from a single ``small"  atom. The quantity $\tau_{Z}^{-2}$ shows different behaviors by increasing cutoff wave number $\Lambda$:  divergence for lin-wQED and convergence for const-wQED; hence a more pronounced memory effect of the radiation field is expected for the latter. Importantly, the collective radiation can be completely established during the Zeno time under the assumption of weak atom-field couplings. More specifically, $\tau_{Z} \gg d/v_{g}$ is guaranteed by $\omega_{0}/\Gamma_{0}\gg \ln (\Lambda/k_{0})$ for lin-wQED and $\omega_{0}/\Gamma_{0}\gg 1$ for const-wQED. We determine the Zeno time for $N$ GAs in Appendix.\ref{A} for completeness.
\begin{figure*}
  \centering
  \hspace{-6.5mm}
  % Requires \usepackage{graphicx}
  \includegraphics[width=18.5cm]{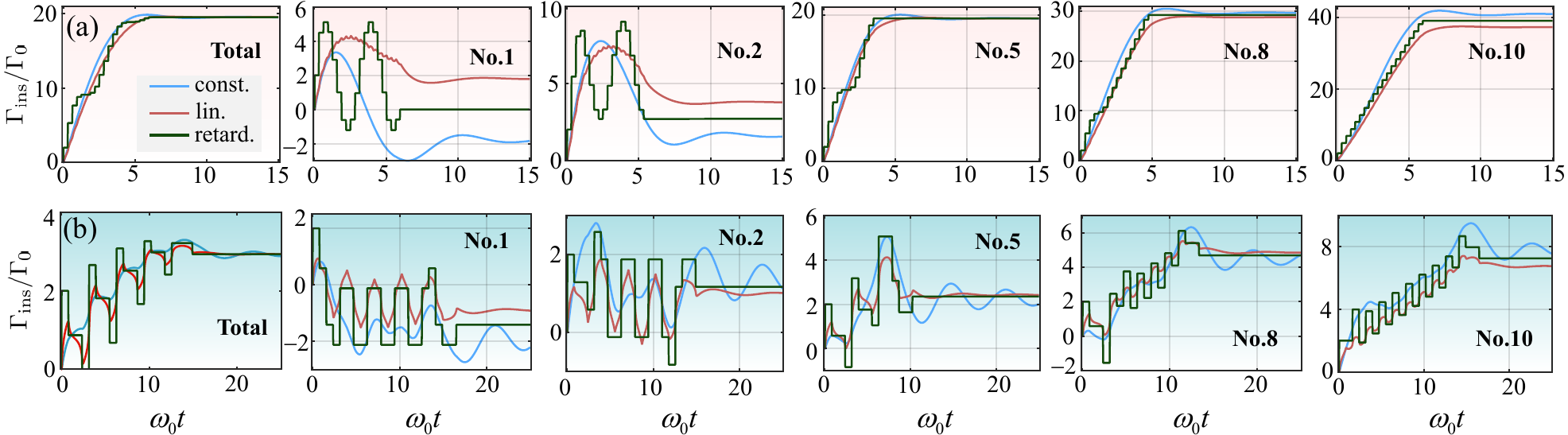}
  \caption{Instantaneous decay rate $\Gamma_{{\rm{ins}}}$ (in units of $\Gamma_{0}$) of (a) $N=10$ separate GAs (background-color: pink) or (b) braided GAs  (background-color: pale green), as a function of scaled time $\omega_{0}t$. The panels labeled by ``Total"  and  ``No.$n$"  represent the decay rate of the whole GAs or $n^{{\rm th}}$ giant atom, respectively. The numerical predictions obtained from const-wQED (blue), lin-wQED (red) or retardation (dark green) have also been marked. The minimal distance $d$ between adjacent coupling points is taken as $d=0.1\pi/k_{0}$ in panel (a) and $d=0.25\pi/k_{0}$ in panel (b).  Other parameters are$M=2$ and $\Gamma_{0}/\omega_{0}=10^{-6}$.}\label{fig3}
\end{figure*}

Since the physical realization of giant-atom structure is facilitated by the superconducting circuits\,\cite{Andersson2015,Kannan2020,PhysRevA.103.023710,PhysRevX.13.021039}, we are now in a position to assess the suitability of the above two setups of waveguide-QED. We first conclude that the lin-wQED setup is wellsuited for giant-atom-based waveguide-QED literature for the following evidences. Under the continuum limit, the conventional 1D transmission line molded by coupled LC loops\,\cite{PhysRevLett.127.233601,PhysRevA.106.063717} enables a coupling architecture with
$g_k\propto \sqrt{\omega _k}\propto \sqrt{\left| \sin k_n\delta _x/2 \right|}\propto \sqrt{|k|}$, where $\delta _x$ is an infinitesimal unit length. For the other model of const-wQED, it is commonly accepted to replace $\omega_{k}$ approximately with atomic resonance frequency $\omega_{0}$ over a wide frequency range, so that $g_k\propto \sqrt{\omega _k}\approx \sqrt{\omega _0}$. Such an interaction mechanism could be implemented by coupling giant atoms to a Josephson photonic crystal waveguide\,\cite{PhysRevLett.126.043602}.

\section{EQUATIONS OF MOTION AND THEIR SOLUTIONS}\label{Sec III}
In this section, we study the dynamical process of spontaneous emission from $N$ GAs into the waveguide, according to the excitations-conservative Hamiltonian $H_{{\rm{eff}}}$.  Assuming that the dynamics of the total GAs plus field system is limited in the single excitation subspace, the state at $t>0$ is
\begin{align}\label{eq6}
|\psi(t)\rangle = \sum_{n=1}^{N}c_{n}(t)\sigma_{n}^{\dagger}|G\rangle+\int_{-\Lambda}^{\Lambda}\frac{dk}{2\pi}\alpha_{k}(t)a_{k}^{\dagger}|G\rangle,
\end{align}
where $c_{n}(t)$ and $\alpha_{k}(t)$ are the amplitudes of probability to find an excitation populated in the $n^{{\rm{th}}}$ giant atom or in propagating mode with wave number  $k$, respectively, at time $t$. $|G\rangle$ is the ground state of the total system, where all of the GAs are in their lower state $\ket{g}$ while the waveguide modes are empty.  Utilizing the Schr$\ddot{o}$dinger equation $i|\dot{\psi}(t)\rangle=H_{{\rm eff}}|\psi(t)\rangle$, the atomic equation of motion (EOM) in the interaction picture is given by
\begin{align}\label{eq7}
\dot{c}_{n}(t)=&-\sum_{n^{\prime}=1}^{N}\sum_{m,m^{\prime}=1}^{M}\int_{-\Lambda}^{\Lambda}\frac{dk}{2\pi}\int_{0}^{t}d\tau|g_{k}^{{\rm JC}}|^{2}c_{n^{\prime}}(\tau)\nonumber\\
& \times e^{i(\omega_{0}-\omega_{k})(t-\tau)}e^{ik(x^{n}_{m}-x^{n^{\prime}}_{m^{\prime}})},
\end{align}
with a mathematical form tracing back the whole historical dynamics of the system. In general, Eq.(\ref{eq7}) has no simple closed form of the analytical description. Here, we present an exact solution  in terms of a convolution integral:
\begin{align}\label{eq8}
c_{n}^{{\rm \frac{const}{lin}}}(t)=c_{n}(0)+\frac{2\Gamma_{0}i}{\pi}\sum_{n^{\prime}=1}^{N}\sum_{m,m^{\prime}=1}^{M}\int_{0}^{t}d\tau c^{{\rm \frac{const}{lin}}}_{n^{\prime}}(\tau)\nonumber\\
\times[\mathcal{K}_{1}(\varphi_{mm'}^{nn'},\phi)-\mathcal{K}_{2}(\varphi_{mm'}^{nn'},\phi)\mp 2\mathcal{K}_{3}(\varphi_{mm'}^{nn'},\phi)],
\end{align}
where $\varphi_{mm'}^{nn'}=k_{0}|x^{n}_{m}-x^{n^{\prime}}_{m^{\prime}}|$ is the acquired field phase and $\phi=\omega_{0}(t-\tau)$, the expressions for the kernels $\mathcal{K}_{1},\mathcal{K}_{2}$ and $\mathcal{K}_{3}$  are given in Appendix.\ref{B}. Note that $c_{n}^{{\rm const}}(t)/c_{n}^{{\rm lin}}(t)$ denotes the dynamical solution for const/lin-wQED [the initial state $c_{n}(t=0)$ does not depend on the specific choices of waveguide setups].

The non-Markovian dynamics in the regime of  non-negligible time delays, can also be derived under several assumptions in Appendix.\ref{B}, and the resulting atomic EOM induced merely by retardation is
\begin{align}\label{eq9}
\dot{c}_{n}(t)\!=\!&-\frac{M}{2}\Gamma_{0}c_{n}(t)\!\!-\!\!\sum_{m\neq m^{\prime}}\frac{\Gamma_{0}}{2}\beta_{n}(\tau_{mm'}^{nn},\varphi_{mm'}^{nn}) \Theta(t\!-\!\tau_{mm'}^{nn})\nonumber\\
&\!\!-\!\sum_{n^{\prime}\neq n}\sum_{m,m^{\prime}=1}^{M}\!\frac{\Gamma_{0}}{2}\beta_{n^{\prime}}(\tau_{mm'}^{nn'},\varphi_{mm'}^{nn'}) \Theta(t\!-\!\tau_{mm'}^{nn'}),
\end{align}
where $\beta_{n}(\tau,\varphi)\equiv c_{n}(t-\tau)e^{i\varphi}$ and $\tau_{mm'}^{nn'}=|x^{n}_{m}-x^{n^{\prime}}_{m^{\prime}}|/v_{g}$ is the time taken by a single photon or phonon traveling between the $m^{{\rm th}}$ coupling point of the $n^{{\rm th}}$ atom and the $m'^{{\rm th}}$ coupling point of the $n'^{{\rm th}}$ atom.

The first term on the right-hand side of Eq. (\ref{eq9}) describes the spontaneous emission processes due to the Markovian dynamics. The second term arising from the nature of nonlocal coupling of a single $M$-legged giant atom, and the remainder describes the atomic relaxation dynamics mediated by the delayed photons released from other GAs.  In contrast to Eq.(\ref{eq7}), the time evolution given by Eq.(\ref{eq9}) predicts that the atomic dynamics response at time $t$ depends only on the state of the system at some specific delay times (this prediction will be labeled as ``retard'' in the following).  Up to now,  the collective emission phenomena of $N$ GAs with arbitrary configurations would be readily simulated based on Eq. (\ref{eq7}) and Eq. (\ref{eq9}).
\begin{figure}
  \centering
  % Requires \usepackage{graphicx}
  \includegraphics[width=8.6cm]{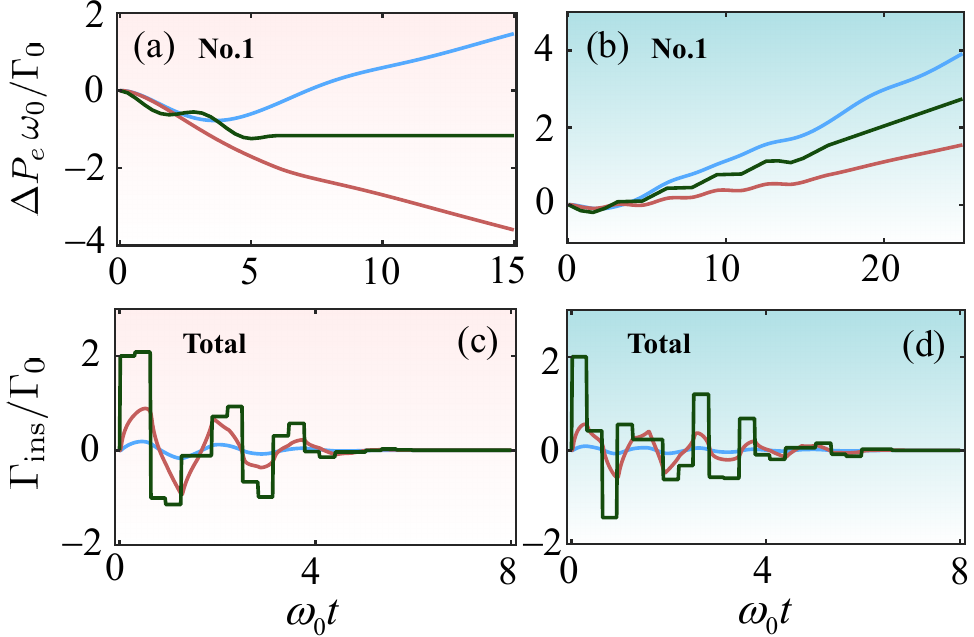}
  \caption{Time evolution of the change on the scaled atomic population $\Delta P_{e}\omega_{0}/\Gamma_{0}$ for atom 1 in the array of separate GAs (a) and braided GAs (b). Subradiant instantaneous decay rates (in units of $\Gamma_{0}$) for separate GAs (c) and braided GAs (d) with fixed $d=0.1\pi/k_{0}$. Other parameters, legend and the notations of background-coloring are the same as Fig.\ref{fig3}.}\label{fig4}
\end{figure}

\section{THE BUILDING OF NON-MARKOVIAN COOPERATIVITY: BEYOND RETARDATION}\label{Sec IV}
In the Zeno regime, the interference properties of the field would be modified drastically due to the joint action of retardation and memory effect of the field. The rich interference between emission from the connected coupling points may enhance or suppress the decay of collective GAs. This phenomenon is famous as super/subradiance\,\cite{PhysRevA.15.2410,RevModPhys.95.015002,science9324}, respectively. Generally speaking, the collective dynamics is sensitive to the specific configurations of GAs, initial states, and the acquired propagating phase between neighboring coupling points. The nonexponential decay of a linear chain of $N$ GAs can be characterized quantitatively by instantaneous decay rate
\begin{align}\label{eq10}
\Gamma_{{\rm ins}}=-\frac{d}{dt}\ln P_{e}(t),
\end{align}
where $P_{e}(t)=\sum\limits_{n}^{N}|c_{n}(t)|^{2}$ for the total decay rate of GAs, and $P_{e}(t)=|c_{n}(t)|^{2}$ for individual decay.
\subsection{Single giant atom: spontaneous emission}
We begin by investigating the spontaneous emission of a single giant atom with $M$ coupling points. As shown in Fig.\ref{fig2}(a), we plot $\Gamma_{{\rm ins}}$ (in units of $\Gamma_{0}$) versus scaled time $\omega_{0}t$ by increasing the number of coupling points from $M=2$ to $M=4$. We show that the decay rates for both of the waveguide QED setups are growing gradually from zero over time until they reach the critical ``radiation burst".  This emerging burst, for instance, is roughly about $5.46\Gamma_{0}$ for $M=2$, after which the decay rates will exhibit oscillating behavior for a while, and eventually be stabilized on the Markovian decay rate
\begin{align}\label{eq11}
\Gamma_{{{\rm Mar}},M}=\Gamma_{0}\frac{\sin^{2} M \varphi/2}{\sin^{2} \varphi /2}
\end{align}
at the long time limit. The non-Markovianity of const- wQED is more obvious than that of lin-wQED according to the anticipation from their Zeno time, i.e., Eq.(\ref{eq4}), and can be enhanced by adding coupling points. Moreover, the influence of ratios $\Gamma_{0}/\omega_{0}$ on the instantaneous decay rates has also been plotted in Fig.\ref{fig2}(b), implying that the establishment of emission is not sensitive to this ratio.

\subsection{Many giant atoms: collective emission}
We proceed by considering $N$ GAs bathed in a common radiation field where the interference behaves constructively. The considered initial state of the system inherits the analog of the timed-Dicke state with the concept originated from the ``small" atoms\,\cite{PhysRevLett.96.010501}. For illustration purposes, we briefly describe the preparation mechanism of the initial state.

The interaction between an incident single photon with wave number $k$ and GAs is
\begin{align}\label{eq12}
\!\!V(t)\!\!=\!\!-i\sum_{n=1}^{N}\sum_{m=1}^{M}g^{{\rm JC}}_{k}\sigma^{\dagger}_{n}a_{k}e^{ikx^{n}_{m}\!-i(\omega_{k}\!-\!\omega_{0})t}\!+\!{\rm{H.c.}},
\end{align}
and the corresponding unitary time evolution operator is $\mathcal{U}(t)=\mathcal{T} e^{-i\int_{0}^{t}V(t')t'}$ where $\mathcal{T}$ is the time ordering operator. When a single resonant photon is incident from the waveguide, the state of the GAs+field system at time $t$ will be evolved into
\begin{align}\label{eq13}
\mathcal{U}(t) a_{k_{0}}^{\dagger}\ket{G}\!\approx \!a_{k_{0}}^{\dagger}\ket{G}\!+\!g^{{\rm JC}}_{k_{0}}t\sum_{n=1}^{N}\sum_{m=1}^{M}\sigma^{\dagger}_{n}e^{ik_{0}x^{n}_{m}}\ket{G},
\end{align}
where $g^{{\rm JC}}_{k_{0}}t$ is assumed to be very small. Therefore, if we fail to probe the photon in the output channel,  then we conclude that the total system has been prepared in
\begin{align}\label{eq14}
\ket{\Psi_{k_{0}}}=(\frac{\Gamma_{0}}{\Gamma_{{{\rm Mar}},M}N})^{1/2}\sum_{n=1}^{N}\sum_{m=1}^{M}\sigma^{\dagger}_{n}e^{ik_{0}x^{n}_{m}}\ket{G}.
\end{align}

The dynamical descriptions in Sec.\,\ref{Sec III} combined with the specific initial state $\ket{\Psi_{k_{0}}}$ enable the whole observation for the building of  collective emission. As shown in Fig.\ref{fig3}, we plot the instantaneous decay rates $\Gamma_{{{\rm ins}}}$ for $N=10$ separate GAs [see Fig.\ref{fig3} (a), background-color: pink] and braided GAs [see Fig.\ref{fig3} (b), background-color: pale green], and the spacing $d$ is taken as $0.1\pi/k_{0}$ and $0.25\pi/k_{0}$, respectively. Here, the curves of  $\Gamma_{{{\rm ins}}}$ for total GAs and individual giant atom have been labeled as ``Total" and ``No.$n$" with $n$ enumerating the emitters, respectively.

The predictions of the total $\Gamma_{{{\rm ins}}}$ from the const-wQED (blue) and lin-wQED (red) setups feature a continuous growth, while the one obtained from only retardation (dark green) gives the steplike increase. These three curves are inconsistent with each other in the early stage of time evolution, but eventually tend to be consistent roughly at $t >10/\omega_{0}$ for separate GAs and longer time for braided GAs.

\begin{figure}
  \centering
  % Requires \usepackage{graphicx}
  \includegraphics[width=8.8cm]{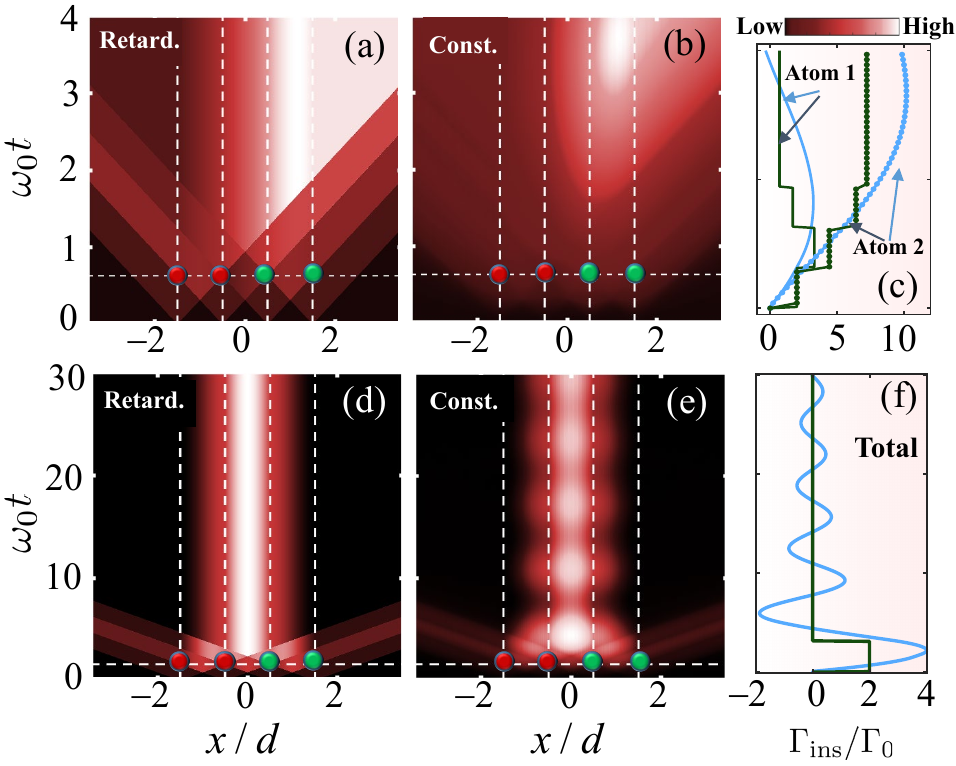}
  \caption{Bosonic field density of emissions from two separate GAs that are prepared initially in (a)-(b) $\ket{\Psi_{k_{0}}}$ and (d)-(e) $\ket{\Psi_{{\rm sub}}}$, respectively. Panels labeled by ``retard"/``const" are plotted according to Eq.(\ref{eq16})/Eq.(\ref{eq17}). Panels (c) and (f) are the corresponding instantaneous decay rates.  We have distinguish the numerical results for atom $1$ (solid lines) and atom $2$ (dotted lines) in (c), and the blue/dark green curves are predictions from const-wQED/retardation. Furthermore, the vertical dashed lines plot the positions of coupling points $x/d=x^{n}_{m}/d$ where $(x^{1}_{1},x^{1}_{2},x^{2}_{1},x^{2}_{2})=(-1.5d,-0.5d,0.5d,1.5d)$; the transverse dashed lines plot the minimal delay time $\omega_{0}t=k_{0}d$. The spacing $d=0.2\pi/k_{0}$ for (a)-(c) and $d=0.5\pi/k_{0}$ for (d)-(f) are used. Other parameters are $M=2,\Gamma_{0}/\omega_{0}=10^{-6}$ and $\Lambda/\omega_{0}=10^{4}$. }\label{fig5}
\end{figure}

For the purpose of further insights of collective behavior in the Zeno regime, we select five GAs, No.$1$, $2$, $5$, $8$, and $10$, and plot their individual decay rate $\Gamma_{{{\rm ins}}}$. We find that the GAs in atomic array could exhibit a wide variety of  decay behaviors. First, the stabilized decay rate of individual GAs tends to increase as they are positioned closer to the right and the scenario will be the opposite when $\ket{\Psi_{k_{0}}}\rightarrow \ket{\Psi_{-k_{0}}}$. Secondly, the three theory frameworks reveal distinct emission dynamics. On the one hand, taking atom 1 (No.$1$) in a separate GAs array as an example, the blue curve shows that the atom quickly enters a long time absorption dynamics after experiencing a short time radiation; the red curve shows that the atom persistently emits excitations into the waveguide while the dark green curve implies two radiation-absorption cycles. Nevertheless, all three curves would be dominated by absorption in the case of braided GAs.  It can also be clearly captured by the change on atomic population $\Delta P_{e}(t)=P_{e}(t)-P_{e}(0)$, as shown in Figs.\ref{fig4}(a) and \ref{fig4}(b). On the other hand, taking atom 10 (No.$10$) as an another example, the dark green curve shows a stable step-like upward trend for the separated GAs. Instead, it shows an oscillating zigzag upward trend for braided GAs, with the upper and lower boundaries of the oscillation being close to blue and red curves periodically. Thus, the received delay signal enable the decay rate that is predicted by the retardation-only picture to exchange faithfully between const-wQED and lin-wQED.  Last but not least, the above three curves agree better at long time for the middle atom (i.e., No.$5$) in atomic array.

We are now in a position to study the subradiant emission of $N$ GAs where the interference between their emissions is destructive.  In the Markovian approximation, the effective Hamiltonian for atomic excitations is available by tracing out the waveguide modes\,\cite{PhysRevX.7.031024,PhysRevLett.123.253601,PhysRevLett.130.023601}, and has the form of
\begin{align}\label{eq15}
\!\!\mathcal{H}_{{\rm eff}}\!=\!\sum_{n,n'=1}^{N}\!(\omega_{0}\delta_{n,n'}\!-\!i\Gamma_{0}\!\!\sum_{m,m'=1}^{N}e^{ik_{0}|x^{n}_{m}\!-\!x^{n'}_{m'}|})\sigma_{n}^{\dagger}\sigma_{n'}.
\end{align}
Numerical diagonalization of $\mathcal{H}_{{\rm eff}}$ in the one-excitation manifold gives $N$ eigenstates. After that the subradiant state $|\Psi_{{\rm sub}}\rangle$ might be found by searching the state with minimal decay rate.

We plot the subradiant decay rates $\Gamma_{{{\rm ins}}}$ for $N=10$ separate GAs and braided GAs in Figs.\ref{fig4} (c) and \ref{fig4}(d), respectively.  It shows that the subradiant emission is established by undergoing several radiation-absorption cycles. The instantaneous decay rates $\Gamma_{{{\rm ins}}}$ oscillate around $0$ with different amplitudes for three predictions. More concretely, the dark green curve predicted from the retardation-only picture has the largest amplitude.  Later discussion shows the opposite case indicating that the const-wQED possesses more strong non-Markovianity in the building of subradiance [see Fig.\,\ref{fig5}(f)]. This pronounced revival of atomic survival probability gives rise to the intriguing phonomenon of oscillating bound states.

\section{CHIRAL EMISSION AND OSCILLATING BOUND STATES}\label{Sec V}
\begin{figure}
  \centering
  % Requires \usepackage{graphicx}
  \includegraphics[width=8.8cm]{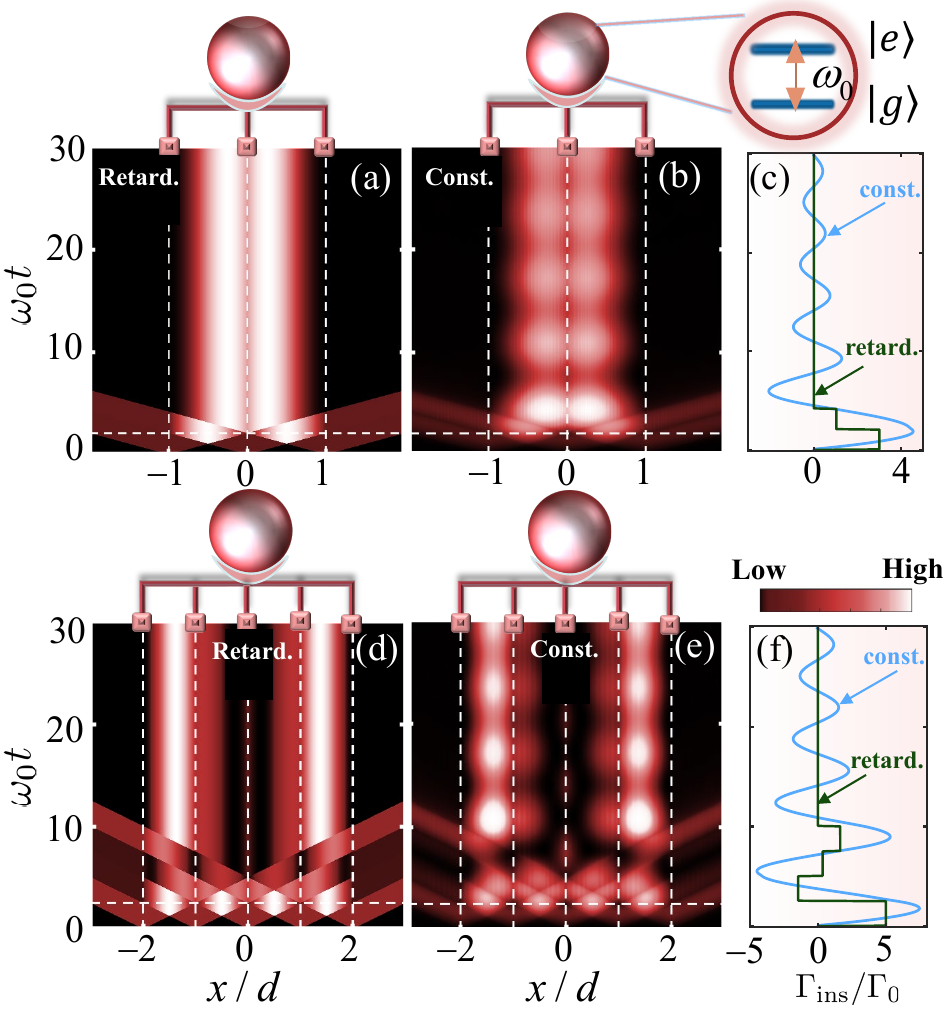}
  \caption{Bosonic field density of emissions from a single giant atom that is prepared initially in its excited state. Upper raw:  intensity distributions for a three-legged giant atom with the numerical predictions are predicted from (a) retardation-only or  (b) cons-wQED. Panel (c) plots the corresponding instantaneous decay rates of spontaneous emission.  The panels in lower raw are the same as that in upper raw but $M=5$, and the distance between adjacent coupling points is $d=0.6641\pi/k_{0}$ for (a)-(c) and $d=0.7985\pi/k_{0}$ for (d)-(f). Other parameters are $\Gamma_{0}/\omega_{0}=10^{-6}$ and $\Lambda/\omega_{0}=10^{6}$. }\label{fig6}
\end{figure}
Until now, we have studied the non-Markovian collective phenomena in a linear chain of $N$ GAs, with the attention mainly focused on atomic dynamics. The asymmetric superradiant behavior in Fig.\ref{fig3} and the significant memory effect of subradiance predicted by const-wQED, inspire us to explore the dynamics of bosonic field. Thus a more complete dynamical description of the atom-field quantum system is expected.

An important figure of merit to quantify the dynamics of the bosonic field in this regard is its bosonic field density\,\cite{PhysRevLett.124.043603,Qiu2023} $I(x,t)\propto \bra{\psi(t)}E^{+}(x,t)E(x,t)\ket{\psi(t)}$, where $E(x,t)$ is the electric field operator. The distribution function $I(x,t)$ describes the probability density at position $x$ and time $t$ to find a single photon for all possible wave vectors $k$. After some algebra presented in Appendix.\,\ref{C}, we find the exact formula of bosonic field density in the retardation-only picture:
\begin{align}\label{eq16}
\!\!\!\!\frac{I(x,t)}{\mathcal{I}_{{\rm retard}}}\!\!=\!\!\big|\sum_{n=1}^{N}\!\sum_{m=1}^{M}\!\sum_{p=\pm}\!\beta_{n}(t_{s}^{p},\!\varphi_{s}^{p})[\Theta(t\!-\!t_{s}^{p})\!-\!\Theta(-t_{s}^{p})]\big|^{2},
\end{align}
where $t_{s}^{\pm}\equiv\pm\frac{x-x^{n}_{m}}{v_{g}}$ and $\varphi_{s}^{p}=k_{0}t_{s}^{p}$. Similarly, the distribution function $I(x,t)$ for const-wQED is given by
\begin{align}\label{eq17}
\frac{I(x,t)}{\mathcal{I}_{{\rm const}}}=\big|\sum_{n=1}^{N}\sum_{m=1}^{M}\int_{0}^{t}d\tau c^{{{\rm const}}}_{n}(\tau)\mathcal{S}_{nm}(t-\tau)\big|^{2},
\end{align}
with the kernel
\begin{align}\label{eq18}
\mathcal{S}_{nm}(t-\tau)=\mathcal{F}_{nm}(\phi,1+\Lambda)-\mathcal{F}_{nm}(\phi,1),
\end{align}
where $\mathcal{F}_{nm}(\phi,x)\equiv \sum\limits_{\zeta=\pm}e^{i(\phi+\phi^{\zeta})}{\rm Csi}^{*}(x\phi^{\zeta})$ and $\phi^{\pm}=\phi\pm k_{0}(x-x^{n}_{m})$ have been introduced for simplicity. Here,  $\mathcal{I}_{{\rm retard}}$ and $\mathcal{I}_{{\rm const}}$ are the normalization constants. One may note that summation terms in Eq.(\ref{eq16}) describe the unidirectional emission from all possible coupling points.  These emissions collide with each other and modify the interference properties of the electromagnetic field, which can be shown from their rich interference patterns.

We plot the bosonic field density $I(x,t)$ for the fields emitted by two separate GAs when atoms are initially prepared in $\ket{\Psi_{k_{0}}}$, as shown in Figs.\ref{fig5}(a) (retard) and \ref{fig5}(b) (const-wQED). The corresponding instantaneous decay rates for atom $1$ (solid lines) and atom $2$ (dotted lines) are plotted in Fig.\ref{fig5}(c). In this case, the atom $2$ decays faster than atom $1$ leading to an  asymmetric distribution of emitted fields. Thus, the interference properties of emissions are strongly modified and eventually exhibit obvious chirality. As a consequence, most of the energy is carried away by the right-moving waveguide modes. Notice that the opposite occurs by replacing $\ket{\Psi_{k_{0}}}\rightarrow \ket{\Psi_{-k_{0}}} $.

We proceed by exploring the intriguing interference behaviors when the GAs build their subradiance. The pronounced revival of atomic excitations in Fig.\,\ref{fig5}(f) at early time implies that the energy is alternately stored  between GAs and the fields.  We plot the bosonic field density $I(x,t)$ for two separate GAs in Figs.\ref{fig5} (d) and (e) according respectively to Eqs. (\ref{eq16}) and  (\ref{eq17}), when atoms are prepared initially in $|\Psi_{{\rm sub}}\rangle$.  The corresponding decay rates for total GAs are shown in panel (f) for comparison purpose. As anticipated, the field intensity predicted by const-wQED is enhanced/weakened during the time window of atomic radiation/absorption stages. Such ``oscillating" behavior of the bound state would vanish when the system dynamics enters completely the regime of exponential decay.

This oscillating behavior for bound states can also arise in the scenario of a single giant atom when the dark state is formed. In order to illustrate it intuitively, we first plot the intensity distributions of emissions emitted from a $M=3$-legged giant atom according to Eqs. (\ref{eq16}) and (\ref{eq17}), as shown respectively in Figs.\ref{fig6}(a) and \ref{fig6}(b). We find the atom periodically emits and absorbs photon or phonon in const-wQED which can also be witnessed in  Fig.\ref{fig6}(c). The similar case occurs for a $5$-legged giant atom [Figs.\ref{fig6}(d)-(f)] with richer interference patterns, where photons or phonons bounce back and forth between certain coupling points. Furthermore, we stress that the discovered oscillating and long-lived bound states in \,\cite{PhysRevResearch.2.043014} are the consequence of coexistence of bound modes in the picture of retardation-only. In our case, the similar phenomenon appears, but the mechanism based on the combination of both non-negligible delay feedback and memory effect of the field only exists in the early stage of spontaneous emission.

\section{CONCLUSION AND OUTLOOKS}\label{Sec VI}
In summary, we have studied the spontaneous radiation of a linear chain of $N$ $M$-legged GAs that are coupled to a 1D waveguide, with arbitrary atomic configurations. We illustrate the collective emission by comparing three completely different theoretical formalisms which give rise to abundant dynamical performance. These predictions enable us to show how GAs build up their collective emission. In particular, the time evolution of atomic population features significant environment memory effect with a highly relaxed temporal resolution. Moreover, we have also investigated the dynamics of the emitted field, characterized by their intensity distribution, to give a more complete perspective for the radiation physics. Interestingly, when prepared in certain superradiant states, the system allows directional emission where most of the energy is carried away either by right or left propagating modes. We also predict a significant oscillating phenomenon in the development of atomic subradiance. Possible applications of our results might be efficient protection of quantum information in the quantum networks with GAs via dissipation engineering\,\cite{Nature1038,Nat660}, or fundamentally useful for theoretical predictions of Zeno physics\,\cite{PhysRevA.81.062131,PhysRevA.106.053709,PhysRevX.13.031009}.

Regarding experimental implementations, we provide a detailed analysis of the experimental feasibility for our scheme. Leveraging the advanced features of the superconducting quantum device, the highly intertwined process of the development for the full collective emission and the transition from quadratic to exponential decay, could be achieved ($\omega _0/2\pi =5.23\mathrm{GHz}, \Gamma _0/2\pi =3.68\mathrm{MHz}$ so that $\tau_{Z}\gg d/v_{g}$ with negligible non-radiative decay $ \Gamma _{\mathrm{nr}}/2\pi =0.03\mathrm{MHz}$)\,\cite{Kannan2020}. Potential experimental challenges encompass the determination of  $\Gamma_{{\rm ins}}(t)$ due to its high requirement in temporal resolution ($\ll \omega_{0}^{-1}$ ), which can be alleviated by considering the dynamics of  $P_{e}(t)$ for atomic array\,\cite{Zhang2023}.

As for future works, it is instructive to explore the many-body signatures of collective emission\,\cite{PhysRevLett.125.263601,PhysRevLett.131.033605,PhysRevX.14.011020,PRXQuantum.5.010344} with GAs in the Zeno regime. In this case, the system dynamics is no longer limited to the single excitation space and interesting multi-particle correlations might emerge. One can also extend the coupling mechanism to that with the higher dimensional photonic or phononic bath\,\cite{PhysRevLett.122.203603,PhysRevResearch.5.023031}.  And interesting physics may be explored by combining the present results with platforms pertaining to topological\,\cite{sciadv0297} or chiral quantum optics\,\cite{Nature21037}.

\section*{ACKNOWLEDGMENTS}
This work is supported by the National Science Fund for Distinguished Young Scholars of China (Grant No. 12425502), the National Key Research and Development Program of China (Grant No. 2021YFA1400700) and the Fundamental Research Funds for the Central Universities (Grant No.2024BRA001). The computation was completed in the HPC Platform of Huazhong University of Science and Technology.

\appendix
\renewcommand\appendixname{APPENDIX}
\section{\uppercase{Calculations of the Zeno time}} \label{A}
In this section, we give the derivations of the Zeno time $\tau_{Z}$ for a single giant atom and $\tau_{Z,N}$ for $N$ GAs, with arbitrary coupling points. The definition of Zeno time is given by the short-time expansion of the atomic survival probability
\begin{align}\label{A1}
\left|\bra{\psi_{0}}e^{-iH_{{\rm JC}}t}\ket{\psi_{0}}\right|^{2}\!=\!1-t^{2}/\tau^{2}_{Z}\!+\!\cdots,
\end{align}
where $\ket{\psi_{0}}$ is the initial state of the emitters+field system.
\subsection*{1. Zeno time for a single giant atom}
We begin by considering an $M$-legged giant atom initially prepared in $\ket{\psi_{0}}=\sigma^{\dagger}\ket{G}$, and then the Zeno time can be obtained easily according to Eq.(\ref{A1})
\begin{align}\label{A2}
\tau_{Z}^{-2}=&\bra{\psi_{0}}H_{{\rm JC}}^{2}\ket{\psi_{0}}-\bra{\psi_{0}}H_{{\rm JC}}\ket{\psi_{0}}^{2}\nonumber\\
&=\sum_{m,m'=1}^{M}\int_{0}^{\Lambda}\frac{dk}{\pi}(g_{k}^{{\rm JC}})^{2}\cos[k(x^{1}_{m}-x^{1}_{m'})].
\end{align}
We then substitute the coupling strength $g_{k}^{{\rm JC}}$ for const-wQED and lin-wQED into Eq.(\ref{A2}), with the obtained results denoted respectively as $\tau_{Z_{{\rm const}}}$ and $\tau_{Z_{{\rm lin}}}$. For the former, we have
\begin{align}\label{A3}
\tau_{Z_{{\rm const}}}^{-2}&=\sum_{m,m'=1}^{M}\int_{0}^{\Lambda}\frac{dk}{\pi}\frac{4\Gamma_{0}v_{g}/2}{(1+k/k_{0})^{2}}\cos(k|x^{1}_{m}-x^{1}_{m'}|)\nonumber\\
&=\frac{2\Gamma_{0}\omega_{0}}{\pi}\big\{M^{2}+2\sum_{n=1}^{M-1}n\varphi(M-n)\times\nonumber\\
&[\sin(n\varphi){\rm{Ci}}(n\varphi z)-\cos(n\varphi){\rm{Si}}(n\varphi z)]|_{1}^{1+\Lambda}\big\},
\end{align}
and for the latter, we have
\begin{align}\label{A4}
\tau_{Z_{{\rm lin}}}^{-2}=&\sum_{m,m'=1}^{M}\int_{0}^{\Lambda}\frac{dk}{\pi}\frac{2\Gamma_{0}v_{g}k/k_{0}}{(1+k/k_{0})^{2}}\cos(k|x^{1}_{m}-x^{1}_{m'}|)\nonumber\\
=&-\tau_{Z_{{\rm const}}}^{-2}+\frac{2\Gamma_{0}\omega_{0}}{\pi}\big\{M\ln(\Lambda+1)+2\sum_{n=1}^{M-1}(M-n)\nonumber\\
&\times[\cos(n\varphi){\rm Ci}(nz\varphi)+\sin(n\varphi){\rm Si}(nz\varphi)]\big|_{1}^{\Lambda+1}\big\}\nonumber\\
\approx &\frac{2\Gamma_{0}\omega_{0}}{\pi}M\ln(\Lambda+1),
\end{align}
where $\varphi=k_{0}d$ and a short hand replacement $\Lambda/k_{0}\rightarrow \Lambda$ has been made.

\subsection*{2. Zeno time for many giant atoms}
Now we consider giant atoms prepared in a so-called timed-Dicke state
\begin{align}\label{A5}
|\psi_{q}\rangle=\mathcal{N}\sum_{n=1}^{N}\sum_{m=1}^{M}e^{iqx^{n}_{m}}\sigma_{n}^{\dagger}|G\rangle,
\end{align}
where $\mathcal{N}=\frac{1}{\sqrt{N}}\frac{\sin(\varphi/2)}{\sin(M\varphi/2)}$ is the normalization coefficient. Thus, the Zeno time $\tau_{Z,N}$ of state $\ket{\psi_{q}}$ can be found by inserting Eq.(\ref{A5}) into Eq.(\ref{A1}):
\begin{align}\label{A6}
\tau_{Z,N}^{-2}=&\bra{\psi_{q}}H_{{\rm JC}}^{2}\ket{\psi_{q}}-\bra{\psi_{q}}H_{{\rm JC}}\ket{\psi_{q}}^{2}\nonumber\\
=&[N\mathcal{N}^{2}\sum_{mm'}e^{iq(x^{1}_{m}-x^{1}_{m'})}\!-\!(N\mathcal{N}^{2}\sum_{mm'}e^{iq(x^{1}_{m}-x^{1}_{m'})})^{2}]\omega_{0}^{2}\nonumber\\
&+\mathcal{N}^{2}\sum_{nn'}^{N}\sum_{mm'}^{M}\int_{-\Lambda}^{\Lambda}\frac{dk}{2\pi}(g_{k}^{{\rm JC}})^{2}e^{i(k-q)(x^{n'}_{m'}-x^{n}_{m})}\nonumber\\
=&F^{2}_{q}\big\{(1-F^{2}_{q})\omega_{0}^{2}+\tau_{Z}^{-2}+\!\frac{1}{N}\sum_{n\neq n'}^{N}\sum_{mm'}^{M}\sum_{jj'}^{M}\nonumber\\
&e^{-iq(x^{n'}_{m'}\!-\!x^{n}_{m})}\int_{-\Lambda}^{\Lambda}\frac{dk}{2\pi}(g_{k}^{{\rm JC}})^{2}e^{ik(x^{n'}_{j'}\!-\!x^{n}_{j})}\big\},
\end{align}
where $F_{q}=\sin^{2}\frac{M}{2}q\varphi/\sin^{2}\frac{1}{2}q\varphi$ and $\tau_{Z}$ is the Zeno time of a single giant atom, i.e.,  Eq.(\ref{A2}). Notice that we have assumed that each giant atom in atomic array is identical, no matter what the configuration of GAs is.
\section{\uppercase{Time evolution of \boldmath$N$ giant atoms}}\label{B}
We are now in a position to discuss the time evolution of the system composed of $N$ GAs coupling to a 1D waveguide.  In the interaction picture, the system Hamiltonian is
\begin{align}\label{B1}
\!H_{{\rm JC}}^{{\rm int}}\!\!=\!\!-i\sum_{n=1}^{N}\sum_{m=1}^{M}\int_{-\Lambda}^{\Lambda}\frac{dk}{2\pi}g_{k}^{{\rm JC}}\sigma_{n}^{\dagger}a_{k}\!e^{i(\omega_{0}\!-\!\omega_{k})t}\!e^{ikx^{n}_{m}}\!+\!{{\rm H.c.}}.
\end{align}
Given the wave function ansatz in the single excitation space
\begin{align}\label{B2}
\ket{\psi(t)}=\sum_{n=1}^{N}c_{n}(t)\sigma_{n}^{\dagger}\ket{G}+\int_{-\Lambda}^{\Lambda}\frac{dk}{2\pi}\alpha_{k}(t)a_{k}^{\dagger}\ket{G},
\end{align}
we can immediately obtain the coupled differential equations of the emitters-field system by applying Schr$\ddot{o}$dinger's equation $i\ket{\dot{\psi}(t)}=H_{{\rm JC}}^{{\rm int}}\ket{\psi(t)}$, which has the form of
\begin{align}
\dot{c}_{n}(t)&=-\sum_{m=1}^{M}\int_{-\Lambda}^{\Lambda}\frac{dk}{2\pi}g_{k}^{{\rm JC}}\alpha_{k}(t)e^{i(\omega_{0}-\omega_{k})t}e^{ikx^{n}_{m}},\label{B3}\\
\dot{\alpha}_{k}(t)&=\sum_{n=1}^{N}\sum_{m=1}^{M}g_{k}^{*{\rm JC}}c_{n}(t)e^{-i(\omega_{0}-\omega_{k})t}e^{-ikx^{n}_{m}}.\label{B4}
\end{align}
The probability amplitudes of bosonic modes can be solved formally as follows
\begin{align}\label{B5}
\!\!\!\alpha_{k}(t)\!=\!\sum_{n=1}^{N}\!\sum_{m=1}^{M}\!g_{k}^{*{\rm JC}}\int_{0}^{t}d\tau c_{n}(\tau)\!e^{-i(\omega_{0}\!-\!\omega_{k})\tau}\!e^{-ikx^{n}_{m}}.
\end{align}
Therefore, the atomic equations of motion can be derived by inserting Eq.(\ref{B5}) into Eq.(\ref{B3}):
\begin{align}\label{B6}
\dot{c}_{n}(t)=&-\sum_{n^{\prime}=1}^{N}\sum_{m,m^{\prime}=1}^{M}\int_{-\Lambda}^{\Lambda}\frac{dk}{2\pi}\int_{0}^{t}d\tau\big|g_{k}^{{\rm JC}}\big|^{2}c_{n^{\prime}}(\tau)\nonumber\\
&\times e^{i(\omega_{0}-\omega_{k})(t-\tau)}e^{ik(x^{n}_{m}-x^{n^{\prime}}_{m^{\prime}})}.
\end{align}
In the following, we will solve numerically Eq.(\ref{B6}) by considering different waveguide-QED setups.
\begin{widetext}
\subsection*{1. Solutions of equations of motion}
The above integro-differential equation \ref{B6} can be transformed to an integral equation
\begin{align}\label{B7}
c_{n}(t)=&c_{n}(0)-\sum_{n^{\prime}=1}^{N}\sum_{m,m^{\prime}=1}^{M}\int_{0}^{t}dt^{\prime}\int_{0}^{\Lambda/k_{0}}\frac{dz}{\pi}\int_{0}^{t^{\prime}}d\tau k_{0}|g^{{\rm JC}}_{k_{0}z}|^{2}c_{n^{\prime}}(\tau)e^{i\omega_{0}(1-z)(t^{\prime}-\tau)}\cos(z\varphi^{nn^{\prime}}_{mm^{\prime}})\nonumber\\
=&c_{n}(0)+\sum_{n^{\prime}=1}^{N}\sum_{m,m^{\prime}=1}^{M}\int_{0}^{t}d\tau c_{n^{\prime}}(\tau)\int_{0}^{\Lambda}dz\frac{\cos(z\varphi_{nn^{\prime},mm^{\prime}})|g^{{\rm JC}}_{k_{0}z}|^{2}k_{0}(1-e^{i\omega_{0}(1-z)(t-\tau)})}{\pi i(1-z)\omega_{0}},
\end{align}
where we have defined $\varphi^{nn'}_{mm^{\prime}}\equiv k_{0}|x^{n}_{m}-x^{n^{\prime}}_{m^{\prime}}|$ for simplicity. We then proceed by plugging the coupling strength of const-wQED $g_{k}=\sqrt{\Gamma_{0}v_{g}/2}$ into Eq.(\ref{B7}), with the obtained solution given by
\begin{align}\label{B8}
c^{{\rm const}}_{n}(t)=&c_{n}(0)+\frac{2\Gamma_{0}i}{\pi}\sum_{n^{\prime}=1}^{N}\sum_{m,m^{\prime}=1}^{M}\int_{0}^{t}d\tau c^{{\rm const}}_{n^{\prime}}(\tau)\int_{0}^{\Lambda}dz\frac{\cos(x\varphi^{nn^{\prime}}_{mm^{\prime}})}{(1+z)^{2}}\frac{1-e^{i\omega_{0}(1-z)(t-\tau)}}{z-1}\nonumber\\
=&c_{n}(0)+\frac{2\Gamma_{0}i}{\pi}\sum_{n^{\prime}=1}^{N}\sum_{m,m^{\prime}=1}^{M}\int_{0}^{t}d\tau c^{{\rm const}}_{n^{\prime}}(\tau)[\mathcal{K}_{1}(\varphi^{nn^{\prime}}_{mm^{\prime}},\phi)-\mathcal{K}_{2}(\varphi^{nn^{\prime}}_{mm^{\prime}},\phi)-2\mathcal{K}_{3}(\varphi^{nn^{\prime}}_{mm^{\prime}},\phi)],
\end{align}
where $\phi=\omega_{0}(t-\tau)$, and the kernels in Eq.(\ref{B8}) given by
\begin{align}\label{B9}
\mathcal{K}_{1}(\varphi^{nn^{\prime}}_{mm^{\prime}},\phi)=&\frac{1}{4}[\cos(\varphi^{nn'}_{mm^{\prime}}){\rm Ci}(z\varphi^{nn^{\prime}}_{mm^{\prime}})-\sin(\varphi^{nn^{\prime}}_{mm^{\prime}}){\rm Si}(z\varphi^{nn^{\prime}}_{mm^{\prime}})\!-\!\frac{e^{i\varphi^{nn^{\prime}}_{mm^{\prime}}}}{2}{\rm Csi}(z\varphi^{nn^{\prime}-}_{mm^{\prime}})\!-\!\frac{e^{-i\varphi^{nn^{\prime}}_{mm^{\prime}}}}{2}{\rm Csi^{*}}(z\varphi^{nn^{\prime}+}_{mm^{\prime}})]\big|_{-1}^{\Lambda-1}\nonumber\\
\mathcal{K}_{2}(\varphi^{nn^{\prime}}_{mm^{\prime}},\phi)=&\frac{1}{4}[\cos(\varphi^{nn^{\prime}}_{mm^{\prime}}){\rm Ci}(z\varphi^{nn^{\prime}}_{mm^{\prime}})+\sin(\varphi^{nn^{\prime}}_{mm^{\prime}}){\rm Si}(z\varphi^{nn^{\prime}}_{mm^{\prime}})-\frac{e^{i(\phi-\varphi^{nn^{\prime}-}_{mm^{\prime}})}}{2}{\rm Csi}(z\varphi^{nn^{\prime}-}_{mm^{\prime}})-\frac{e^{i(\phi+\varphi^{nn^{\prime}+}_{mm^{\prime}})}}{2}{\rm Csi^{*}}(z\varphi^{nn^{\prime}+}_{mm^{\prime}})]\big|_{1}^{\Lambda+1}\nonumber\\
\mathcal{K}_{3}(\varphi^{nn^{\prime}}_{mm^{\prime}},\phi)=&\frac{1}{4}\frac{\cos(\Lambda\varphi^{nn^{\prime}}_{mm^{\prime}})}{\Lambda+1}[e^{-i(\Lambda-1)\phi}-1]+1-e^{i\phi}-\big\{\varphi^{nn^{\prime}}_{mm^{\prime}}[\cos(\varphi^{nn^{\prime}}_{mm^{\prime}}){\rm Si}(z\varphi^{nn^{\prime}}_{mm^{\prime}})-\sin(\varphi^{nn^{\prime}}_{mm^{\prime}}){\rm Ci}(z\varphi^{nn^{\prime}}_{mm^{\prime}})]\nonumber\\
&+\frac{i}{2}\varphi^{nn^{\prime}-}_{mm^{\prime}}e^{i(2\ensuremath{\phi-\varphi^{nn^{\prime}}_{mm^{\prime}}})}{\rm Csi}(z\varphi^{nn^{\prime}-}_{mm^{\prime}})-\frac{i}{2}\varphi^{nn^{\prime}+}_{mm^{\prime}}e^{i(2\ensuremath{\phi+\varphi^{nn^{\prime}}_{mm^{\prime}}})}{\rm Csi}(z\varphi^{nn^{\prime}+}_{mm^{\prime}})\big\}\big|_{1}^{\Lambda+1},
\end{align}
where $\varphi^{nn^{\prime}\pm}_{mm^{\prime}}\equiv\varphi^{nn^{\prime}}_{mm^{\prime}}\pm \phi$ is introduced. Following similar procedures, we can easily obtain the time evolution of atomic population for lin-wQED, which is given by
\begin{align}\label{B10}
c^{{\rm lin}}_{n}(t)=c_{n}(0)+\frac{2\Gamma_{0}i}{\pi}\sum_{n^{\prime}=1}^{N}\sum_{m,m^{\prime}=1}^{M}\int_{0}^{t}d\tau c^{{\rm lin}}_{n^{\prime}}(\tau)[\mathcal{K}_{1}(\varphi^{nn^{\prime}}_{mm^{\prime}},\phi)-\mathcal{K}_{2}(\varphi^{nn^{\prime}}_{mm^{\prime}},\phi)+2\mathcal{K}_{3}(\varphi^{nn^{\prime}}_{mm^{\prime}},\phi)].
\end{align}
As the dynamics controlled by Eq.(\ref{B1}) conserves the total number of excitations, thus the normalization of  atomic probability amplitudes obtained in  Eq.(\ref{B8}) and  Eq.(\ref{B10}) should be done by making a transformation $c_{n}(t)\rightarrow c_{n}(t)/\sqrt{|c_{n}(t)|^{2}+N_{B}(t)}$. Here, $N_{B}(t)=\int_{-\Lambda}^{\Lambda}\frac{dk}{2\pi}|\alpha_{k}(t)|^{2}$ is the number of excitations in the 1D waveguide. After some algebra, the expression of  $N_{B}(t)$ has the form of
\begin{align}\label{B11}
N_{B}(t)\!=\!\sum_{n,n^{\prime}=1}^{N}\sum_{m,m^{\prime}=1}^{M}\int_{0}^{t}d\tau\int_{0}^{t}d\tau^{\prime}c_{n}(\tau)c_{n^{\prime}}(\tau^{\prime})B_{nn^{\prime};mm^{\prime}}(\tau\!-\!\tau^{\prime}),
\end{align}
where the kernel $B_{nn^{\prime};mm^{\prime}}$ is given by
\begin{align}\label{B12}
B^{{\rm const}}_{nn^{\prime};mm^{\prime}}(\tau\!-\!\tau^{\prime})=\frac{2\Gamma_{0}\omega_{0}}{\pi}e^{-i\phi}\big[-\frac{\cos(z\varphi^{nn^{\prime}}_{mm^{\prime}})}{1+z}e^{i\phi z}\big|_{0}^{\Lambda}
+\sum_{s=\pm}\frac{i}{2}\phi_{s}e^{-i\phi_{s}}{\rm Csi}(z \phi_{s})\big|_{1}^{\Lambda+1}\big]
\end{align}
for const-wQED, and
\begin{align}\label{B13}
B^{{\rm lin}}_{nn^{\prime};mm^{\prime}}(\tau\!-\!\tau^{\prime})=\frac{\Gamma_{0}\omega_{0}}{\pi}e^{-i\phi}\sum_{s=\pm}e^{-i\phi_{s}}{\rm Csi}(z \phi_{s})\big|_{1}^{\Lambda+1}
-B_{nn^{\prime};mm^{\prime}}^{{\rm const}}(\tau-\tau^{\prime})
\end{align}
for lin-wQED, where $\phi_{\pm}=\phi\pm\varphi^{nn^{\prime}}_{mm^{\prime}}$ is introduced. Up to now, the exact solutions for atomic time evolution are found for both the waveguide setups, as shown in Eq.(\ref{eq8}) in the main text.
\subsection*{2. Non-Markovian dynamics in the retardation-only picture}
In order to obtain the equations of motion considering only the retardation, one can rewrite Eq.(\ref{B6}) in the limit of $\Lambda\rightarrow\infty$ as follows
\begin{align}\label{B14}
\dot{c}_{n}(t)=&-\sum_{n^{\prime}=1}^{N}\sum_{m,m^{\prime}=1}^{M}\int_{-\infty}^{\infty}\frac{dk}{2\pi}\int_{0}^{t}d\tau|g_{k}^{{\rm JC}}|^{2}c_{n^{\prime}}(\tau)e^{i(\omega_{0}-\omega_{k})(t-\tau)}e^{ik(x^{n}_{m}-x^{n^{\prime}}_{m^{\prime}})}\nonumber\\
=&-\sum_{n^{\prime}=1}^{N}\sum_{m,m^{\prime}=1}^{M}\int_{0}^{\infty}\frac{d\omega_{k}}{2\pi v_{g}}|g_{k}^{{\rm JC}}|^{2}\int_{0}^{t}d\tau c_{n^{\prime}}(\tau)\big[e^{i\omega_{0}(t-\tau)}e^{i\omega_{k}\frac{(x^{n}_{m}-x^{n^{\prime}}_{m^{\prime}})}{v_{g}}-(t-\tau)}+e^{i\omega_{0}(t-\tau)}e^{i\omega_{k}\frac{(-x^{n}_{m}+x^{n^{\prime}}_{m^{\prime}})}{v_{g}}-(t-\tau)}\big]\nonumber\\
\approx&-\frac{|g_{k_{0}}^{{\rm JC}}|^{2}}{v_{g}}[2M\int_{0}^{t}d\tau c_{n}(\tau)\delta(t-\tau)+\sum_{m\neq m^{\prime}}\int_{0}^{t}d\tau c_{n}(\tau)\delta(t-\frac{|x^{n}_{m}-x^{n}_{m^{\prime}}|}{v_{g}}-\tau)e^{ik_{0}|x^{n}_{nm}-x^{n}_{m^{\prime}}|}]\nonumber\\
&-\sum_{n^{\prime}\neq n}^{N}\sum_{m,m^{\prime}=1}^{M}\frac{\Gamma_{0}}{2}c_{n^{\prime}}(t-\frac{|x^{n}_{m}-x^{n^{\prime}}_{m^{\prime}}|}{v_{g}})e^{ik_{0}|x^{n}_{m}-x^{n^{\prime}}_{m^{\prime}}|}\Theta(t-\frac{|x^{n}_{m}-x^{n^{\prime}}_{m^{\prime}}|}{v_{g}})\nonumber\\
\!\!=\!&-\frac{M}{2}\Gamma_{0}c_{n}(t)\!-\!\sum_{m\neq m^{\prime}}\frac{\Gamma_{0}}{2}c_{n}(t\!-\!\frac{\varphi^{nn}_{mm^{\prime}}}{\omega_{0}})e^{i\varphi^{nn}_{mm^{\prime}}}\Theta(t\!-\!\frac{\varphi^{nn}_{mm^{\prime}}}{\omega_{0}})
\!-\!\sum_{n^{\prime}\neq n}^{N}\sum_{m,m^{\prime}=1}^{M}\frac{\Gamma_{0}}{2}c_{n^{\prime}}(t\!-\!\frac{\varphi^{nn'}_{mm^{\prime}}}{\omega_{0}})e^{i\varphi^{nn'}_{mm^{\prime}}}\Theta(t\!-\!\frac{\varphi^{nn'}_{mm^{\prime}}}{\omega_{0}})\nonumber\\
=&-\frac{M}{2}\Gamma_{0}c_{n}(t)\!-\!\sum_{m\neq m^{\prime}}\frac{\Gamma_{0}}{2}\beta_{n}(\tau_{mm'}^{nn},\varphi_{mm'}^{nn}) \Theta(t\!-\!\tau_{mm'}^{nn})
\!-\!\sum_{n^{\prime}\neq n}\sum_{m,m^{\prime}=1}^{M}\frac{\Gamma_{0}}{2}\beta_{n^{\prime}}(\tau_{mm'}^{nn'},\varphi_{mm'}^{nn'}) \Theta(t\!-\!\tau_{mm'}^{nn'}),
\end{align}
which is exactly the Eq.(\ref{eq9}) in the main text, and $\Theta (\bullet)$ is the Heaviside step function. Such a simplified equation of motion for GAs indicates that the time evolution of atom $n$ at time $t$ is decided only by the historical dynamics at certain instants $\tau_{mm'}^{nn'}$. As a consequence, it shows great discrepancy from that predicted from const-wQED and lin-wQED.
\end{widetext}
\section{\uppercase{Detailed calculations of field intensity distribution}}\label{C}
We now turn to focus on the dynamical properties of the bosonic field emitted from the coupled GAs, which can be characterized by its field intensity distribution
\begin{align}\label{C1}
I(x,t)=\frac{\epsilon_{0}v_{g}}{2}\bra{\psi(t)}E^{\dagger}(x,t)E(x,t)\ket{\psi(t)},
\end{align}
where $E(x,t)$ is the electric field operator at position $x$ and time $t$ with a general form
\begin{align}\label{C2}
\!\!E(x,t)\!=\!\int_{0}^{\Lambda}dk\mathcal{E}_{k}[e^{ikx}a_{R}(k)\!+\!e^{-ikx}a_{L}(k)]e^{-i\omega_{k}t}.
\end{align}
 Here, $a_{R/L}(k)$ is the annihilation operator of the right-/left-moving mode with frequency $\omega_{k}$ and $\mathcal{E}_{k}$ is assumed to be a constant $\mathcal{E}_{k_{0}}$ for all $k$. Combined with the Eq.(\ref{B5}) and  Eq.(\ref{C2}), the field density distribution $I(x,t)$ for const-wQED can be calculated as follows
\begin{align}\label{C3}
I(x,t)=&\frac{\epsilon_{0}v_{g}|\mathcal{E}_{k_{0}}\big|^{2}}{2}|\langle G|\int_{0}^{\Lambda}dk[e^{ikx}a_{R}(k)+e^{-ikx}a_{L}(k)]\nonumber\\
&\times e^{-i\omega_{k}t}|\psi(t)\rangle\big|^{2}\nonumber\\
=&8\Gamma_{0}v_{g}I_{0}\big|\sum_{n=1}^{N}\sum_{m=1}^{M}\int_{0}^{t}d\tau\int_{0}^{\Lambda}dk\frac{1}{1+k/k_{0}}c_{n}(\tau)\nonumber\\
&\cos[k(x-x_{m}^{n})]e^{-i(\omega_{k}-\omega_{0})(t-\tau)}\big|^{2}\nonumber\\
=&2\Gamma_{0}v_{g}I_{0}\big|\sum_{n=1}^{N}\sum_{m=1}^{M}\int_{0}^{t}d\tau c_{n}(\tau)\int_{1}^{1+\Lambda/k_{0}}dz\frac{k_{0}}{z}\nonumber\\
\!&\times[e^{i(r_{nm}-\phi)z}e^{i(2\phi-r_{nm})}\!+\!e^{-i(r_{nm}\!+\!\phi)z}e^{i(2\phi\!+\!r_{nm})}]\big|^{2}\nonumber\\\!
\!\!=&2\Gamma_{0}v_{g}I_{0}|\sum_{n=1}^{N}\sum_{m=1}^{M}\int_{0}^{t}d\tau c_{n}(\tau)\!S_{nm}(t\!-\!\tau)|^{2}\!,\!\!
\end{align}
which is exactly Eq.(\ref{eq17}) in the main text, where $I_{0}=\frac{\epsilon_{0}v_{g}|\mathcal{E}_{k_{0}}|^{2}}{2}, r_{nm}=k_{0}(x-x_{m}^{n})$ and the atomic probability amplitude $ c_{n}(\tau)$ is given by \,Eq.(\ref{B8}). The kernel $S_{nm}(t-\tau)$ in\,Eq.(\ref{C3}) reads
\begin{align}\label{C4}
\mathcal{S}_{nm}(t-\tau)=\mathcal{F}_{nm}(\phi,1+\Lambda)-\mathcal{F}_{nm}(\phi,1).
\end{align}
Here, we have introduced the notation $\mathcal{F}_{nm}(\phi,x)\equiv\sum\limits_{\zeta=\pm}e^{i(\phi+\phi_{\zeta})}{\rm Csi}^{*}(x\phi_{\zeta})$ for simplicity.

We proceed by calculating the field intensity of the emitted field in the retardation-only picture. In the limit of large enough $\Lambda\rightarrow\infty$, we have
\begin{widetext}
\begin{align}\label{C5}
I(x,t)=&\frac{\epsilon_{0}v_{g}|\mathcal{E}_{k_{0}}\big|^{2}}{2}|\langle G|\int_{0}^{\infty}dk[e^{ikx}a_{R}(k)+e^{-ikx}a_{L}(k)]
\times e^{-i\omega_{k}t}|\psi(t)\rangle\big|^{2}\nonumber\\
=&\frac{I_{0}\Gamma_{0}}{2v_{g}}\Big|\int_{-\infty}^{\infty}dz\int_{0}^{t}d\tau\sum_{n=1}^{N}\sum_{m=1}^{M}c_{n}(\tau)[e^{iz(\frac{x-x_{m}^{n}}{v_{g}}+\tau-t)}
\times e^{i\omega_{0}(\frac{x-x_{m}^{n}}{v_{g}}-t)}+e^{-iz(\frac{x-x_{m}^{n}}{v_{g}}-\tau+t)}e^{-i\omega_{0}(\frac{x-x_{m}^{n}}{v_{g}}+t)}]\Big|^{2}\nonumber\\
=&\frac{2\pi^{2}I_{0}\Gamma_{0}}{v_{g}}\Big|\sum_{n=1}^{N}\sum_{m=1}^{M}\big\{c_{n}(t\!-\!x_{m}^{n}/v_{g}\!+\!x/v_{g})e^{i\omega_{0}(x_{m}^{n}\!-\!x)/v_{g}}
[\Theta(-\frac{x_{m}^{n}\!-\!x}{v_{g}})\!-\!\Theta(t\!-\!\frac{x_{m}^{n}\!-\!x}{v_{g}})]\nonumber\\
&+c_{n}(t\!+\!x_{m}^{n}/v_{g}\!-\!x/v_{g})e^{-i\omega_{0}(x_{m}^{n}-x)/v_{g}}[\Theta(\frac{x_{m}^{n}-x}{v_{g}})-\Theta(t+\frac{x_{m}^{n}-x}{v_{g}})]\big\}\Big|^{2}\nonumber\\
=&\frac{2\pi^{2}I_{0}\Gamma_{0}}{v_{g}}|\sum_{n=1}^{N}\sum_{m=1}^{M}\sum_{p=\pm}c_{n}(t-p\frac{x-x_{m}^{n}}{v_{g}})e^{i\omega_{0}p\frac{x-x_{m}^{n}}{v_{g}}}
\times[\Theta(t-p\frac{x-x_{m}^{n}}{v_{g}})-\Theta(-p\frac{x-x_{m}^{n}}{v_{g}})]|^{2},
\end{align}
%
%%
%\begin{align}\label{C5}
%I(x,t)=&\frac{\epsilon_{0}v_{g}|\mathcal{E}_{k_{0}}\big|^{2}}{2}|\langle G|\int_{0}^{\infty}dk[e^{ikx}a_{R}(k)+e^{-ikx}a_{L}(k)]\nonumber\\
%&\times e^{-i\omega_{k}t}|\psi(t)\rangle\big|^{2}\nonumber\\
%=&\frac{I_{0}\Gamma_{0}}{2v_{g}}\Big|\int_{-\infty}^{\infty}dz\int_{0}^{t}d\tau\sum_{n=1}^{N}\sum_{m=1}^{M}c_{n}(\tau)[e^{iz(\frac{x-x_{m}^{n}}{v_{g}}+\tau-t)}\nonumber\\
%&\times e^{i\omega_{0}(\frac{x-x_{m}^{n}}{v_{g}}-t)}+e^{-iz(\frac{x-x_{m}^{n}}{v_{g}}-\tau+t)}e^{-i\omega_{0}(\frac{x-x_{m}^{n}}{v_{g}}+t)}]\Big|^{2}\nonumber\\
%\!=&\frac{2\pi^{2}I_{0}\Gamma_{0}}{v_{g}}\Big|\sum_{n=1}^{N}\sum_{m=1}^{M}\big\{c_{n}(t\!-\!x_{m}^{n}/v_{g}\!+\!x/v_{g})e^{i\omega_{0}(x_{m}^{n}\!-\!x)/v_{g}}\nonumber\\
%\!&\![\Theta(-\frac{x_{m}^{n}\!-\!x}{v_{g}})\!-\!\Theta(t\!-\!\frac{x_{m}^{n}\!-\!x}{v_{g}})]\!+\!c_{n}(t\!+\!x_{m}^{n}/v_{g}\!-\!x/v_{g})\nonumber\\
%&\times e^{-i\omega_{0}(x_{m}^{n}-x)/v_{g}}[\Theta(\frac{x_{m}^{n}-x}{v_{g}})-\Theta(t+\frac{x_{m}^{n}-x}{v_{g}})]\big\}\Big|^{2}\nonumber\\
%=&\frac{2\pi^{2}I_{0}\Gamma_{0}}{v_{g}}|\sum_{n=1}^{N}\sum_{m=1}^{M}\sum_{p=\pm}c_{n}(t-p\frac{x-x_{m}^{n}}{v_{g}})e^{i\omega_{0}p\frac{x-x_{m}^{n}}{v_{g}}}\nonumber\\
%&\times[\Theta(t-p\frac{x-x_{m}^{n}}{v_{g}})-\Theta(-p\frac{x-x_{m}^{n}}{v_{g}})]|^{2},
%\end{align}
%%
\end{widetext}
which is exactly Eq.(\ref{eq16}) in the main text. Therefore, the calculations of the field intensity distribution for const-wQED and the one with only retardation are completed.
%\bibliography{refer}
%merlin.mbs apsrev4-1.bst 2010-07-25 4.21a (PWD, AO, DPC) hacked
%Control: key (0)
%Control: author (72) initials jnrlst
%Control: editor formatted (1) identically to author
%Control: production of article title (-1) disabled
%Control: page (0) single
%Control: year (1) truncated
%Control: production of eprint (0) enabled
%

\end{document}